\documentclass{PoS}
\newcommand{\integral}{{\it INTEGRAL}}
\newcommand{\rxte}{{\it RXTE}}
\newcommand{\beppo}{{\it BeppoSAX}}
\newcommand{\mysou}{{GX~339$-$4}}
\newcommand{\cyg}{{Cyg~X-1}}
\def\ergcms{erg cm$^{-2}$ s$^{-1}$ }

\bibliography{/data/Proceeding/Microquasar_como/Pos.bst}
\title{Spectral variability modes of GX~339$-$4 in a hard-to-soft state transition}

\ShortTitle{GX 339$-$4 variability modes}

\author{\speaker{Melania Del Santo}\\
        INAF/IASF-Roma, Italy\\
        E-mail: \email{melania.delsanto@iasf-roma.inaf.it}}

\author{Julien Malzac\\
        CNRS/CESR Toulouse, France}

\author{Pietro Ubertini\\
        INAF/IASF-Roma, Italy}

\author{Tomaso Belloni\\
        INAF/OAB Merate, Italy}

\abstract{We report on \integral~ observations performed during the 2004 outburst of the bright black
hole transient \mysou. We analysed IBIS and JEM-X public data starting on 9th August
and lasting about one month.
During this period \mysou~ showed spectral state transitions.
In order to seek for variability patterns, a principal component analysis (PCA)
has been used.}

\FullConference{VI Microquasar Workshop: Microquasars and Beyond\\
		September 18-22 2006\\
		Societ\`a del Casino, Como, Italy}

\begin{document}

\begin{figure}[t!]
\centering
\includegraphics[height=9cm,angle=90]{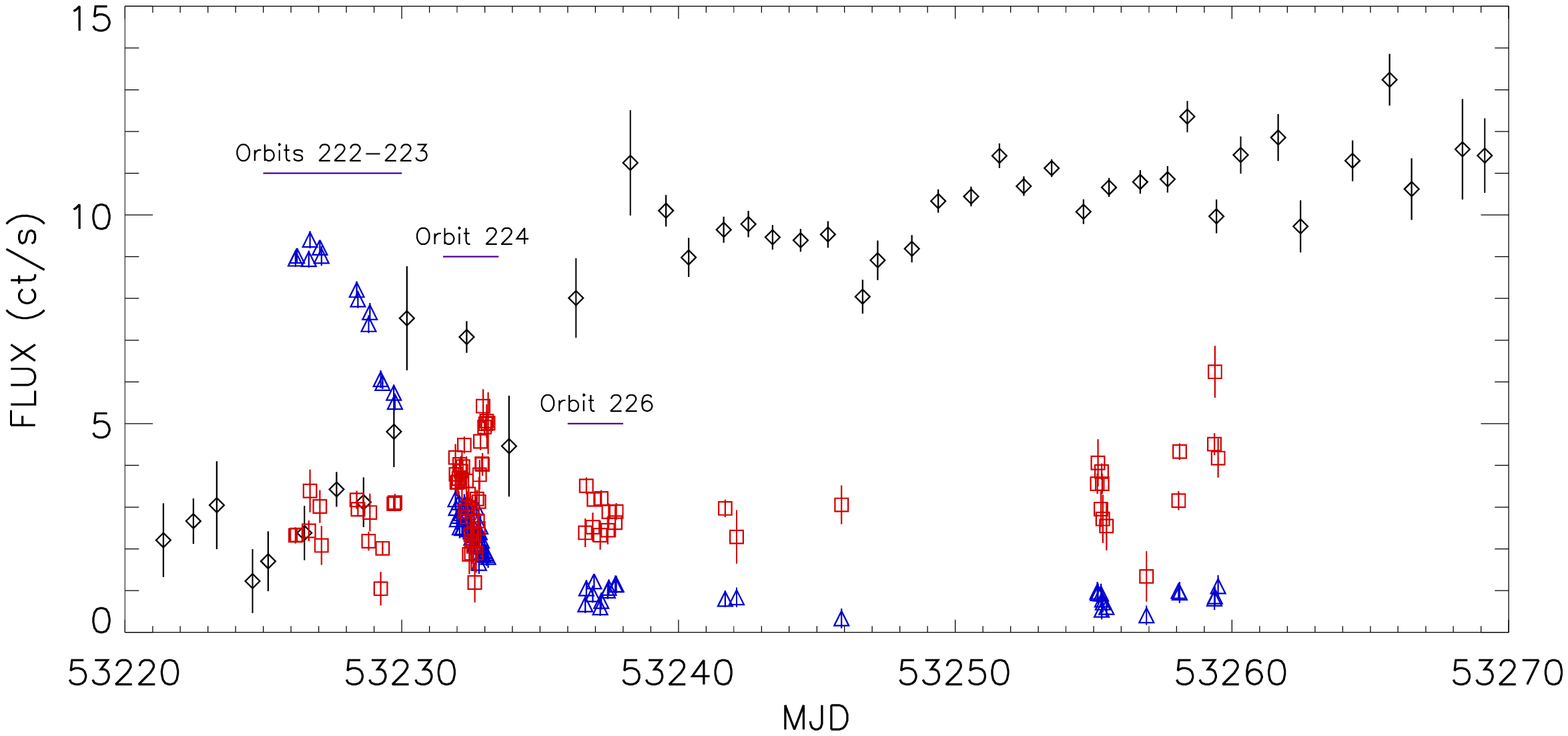}
\includegraphics[height=9cm,angle=90]{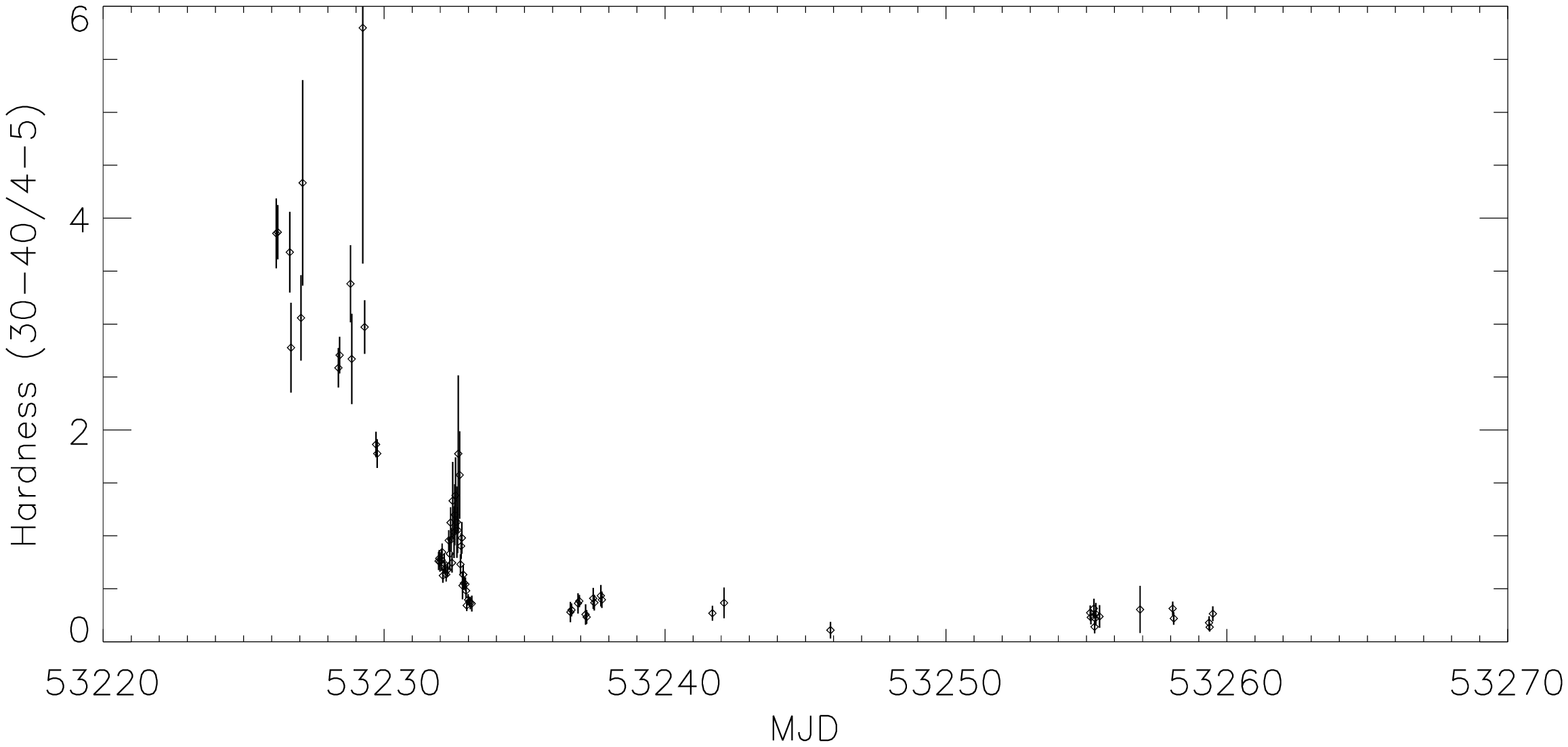}
\caption{\mysou~ count rate in 30-40 keV, 4-5 keV and 1.5-3 keV with IBIS/ISGRI (blue triangles), 
JEM-X (red squares) and ASM (black diamonds), respectively ({\it top}). IBIS/ISGRI to JEM-X hardness ratio ({\it bottom}). \label{fig:lc}}
\end{figure}

\section{Introduction}
Classified as black-hole candidate \cite{zdz98}, \mysou~ is a transient source 
spending long periods in outburst. Before the launch of \rxte~ the source has been prevalently 
observed in the Low/Hard (LH) spectral state, even though several transitions to softer states 
were reported \cite{belloni05a}. Thereafter \mysou~ remained bright and mostly in the LH state until 1999 when it went into 
quiescence. After the quiescent state which was observed by \beppo~ \cite{kong00}, \cite{corbel03},
\mysou~ showed two new outbursts: in 2002/2003 \cite{smith02}, \cite{nespoli03}, \cite{belloni05a} and   
in 2004 after one year in quiescence \cite{belloni04}, \cite{buxton04}.  
The long-term variability of \mysou~ and a theoretical 
interpretation referred to the time range 1987-2004 (only the first LH period for the latter) 
is extensively described in Zdziarski et al. (2004). These authors studied 
in detail the hysteresis-like behaviour, especially the strong dependence of the flux of the 
hard-to-soft transition on the preceding behaviour of the hard/quiescent state. 
The 2004 outburst started on February and occurred at luminosity levels lower than the 
previous one. Joint \rxte/\integral~ observations performed on 2004 August 14$^{th}$-16$^{th}$ are presented 
in Belloni et al. (2006). These authors report on the first determination for a BHC of the changes 
of the broad-band X-ray spectrum across the High Intermediate State (HIMS)-Soft Intermediate State (SIMS) 
fast transition \cite{belloni05b}. 

In order to seek for a variability pattern of \cyg, Malzac et al. (2006) used a Principal Component Analysis (PCA) 
during an intermediate state of that source. 
These authors found that the spectral variability occurred 
through two dependent modes: the first consisted in changes in the overall luminosity on time scale 
of hours with almost constant spectrum (responsible for 68\% of the variance); 
the second mode consisted in a pivoting of the spectrum around 10 keV (27\% of the variance). 

In this work we present \mysou~ spectral analysis of \integral~ data collected in the two adjacent periods 
to the one used by Belloni et al. (2006), as well as a PCA on the spectral transition period.

\begin{figure*}[t!]
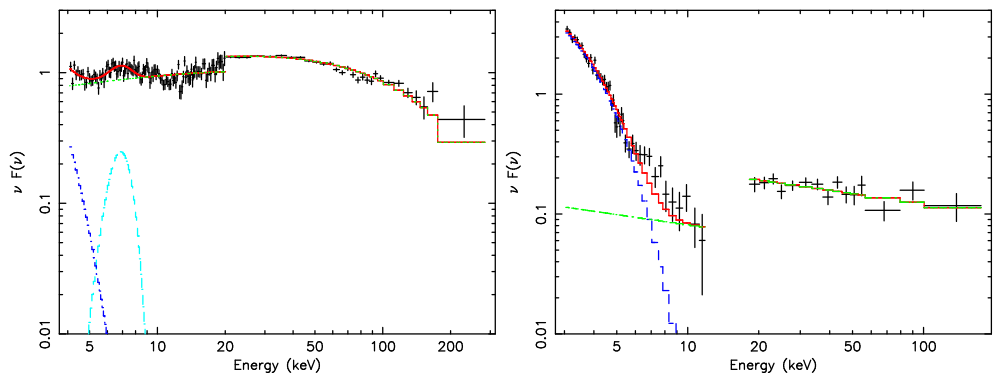

\centering
\includegraphics[height=6.5cm,angle=-90]{rev222_223_nufnu.ps}
\includegraphics[height=6.5cm,angle=-90]{rev226_nufnu.ps}
\caption{Energy spectrum (\ergcms), total model and single components of the HIMS ({\it left}) and 
SIMS ({\it right}) spectra. The JEM-X and IBIS/ISGRI spectra shown are not normalised.
 \label{fig:spec}}
\end{figure*}

\section{Observations and data analysis}
We used \integral~data collected during the period 9 August-10 September 2004, which covered range 222$\div$233
of the satellite orbits.
In particular, we selected a data sub-set with all pointings common to JEM-X and IBIS telescopes 
for a total of 76 Science Windows (SCWs), lasting 1800-3600 seconds each.
\integral~ data analysis has been performed with OSA v. 5.1. 

Light curves for each pointing were extracted in 15 bins\footnote{Namely: 3-4, 4-5, 5-6, 6-7, 7-9, 9-11,
11-13, 13-15, 15-20, 20-30, 30-40, 40-50, 50-70, 70-100, 100-200 keV}. 
The IBIS/ISGRI and the JEM-X temporal behaviours are shown in Fig. \ref{fig:lc}. 
The Belloni et al. (2006) data correspond to orbit 224, during which the transition occurred.

Spectral fitting of two average spectra ({\it spec1} and {\it spec2}),
collected within orbits 222-223 and 226, respectively, is presented.
We used simple models as multicolor disc and cut-off or simple power laws in XSPEC v. 11.3.1. 
The single parameters uncertainties have been calculated at the 90\% confidence level
and the cross-calibration constant is 0.7 with 0.2 of uncertainty ($const=1$ for IBIS/ISGRI). 

\subsection{The principal Component Analysis}
Principal Component Analysis (PCA) is a powerful tool for multivariate data analysis used for a broad range 
of applications in natural as well as social science \cite{kendall80}. 
The main use of PCA is to reduce the dimensionality of a data set while keeping
as much informations as possible.
PCA transforms a number of (possibly) correlated variable in a smaller number of uncorrelated variables
called principal components. 

Details on the used method are described in Malzac et al. (2006).  
In our case, {\it p} = 76 spectra (one for each SCW) binned into {\it n}~=~15 bins corresponding to
energies {\it E$_{1}$}, {\it E$_{2}$}, ..., {\it E$_{n}$}.

\begin{figure*}[t!]
\centering
\includegraphics[height=9.cm, width=4.6cm]{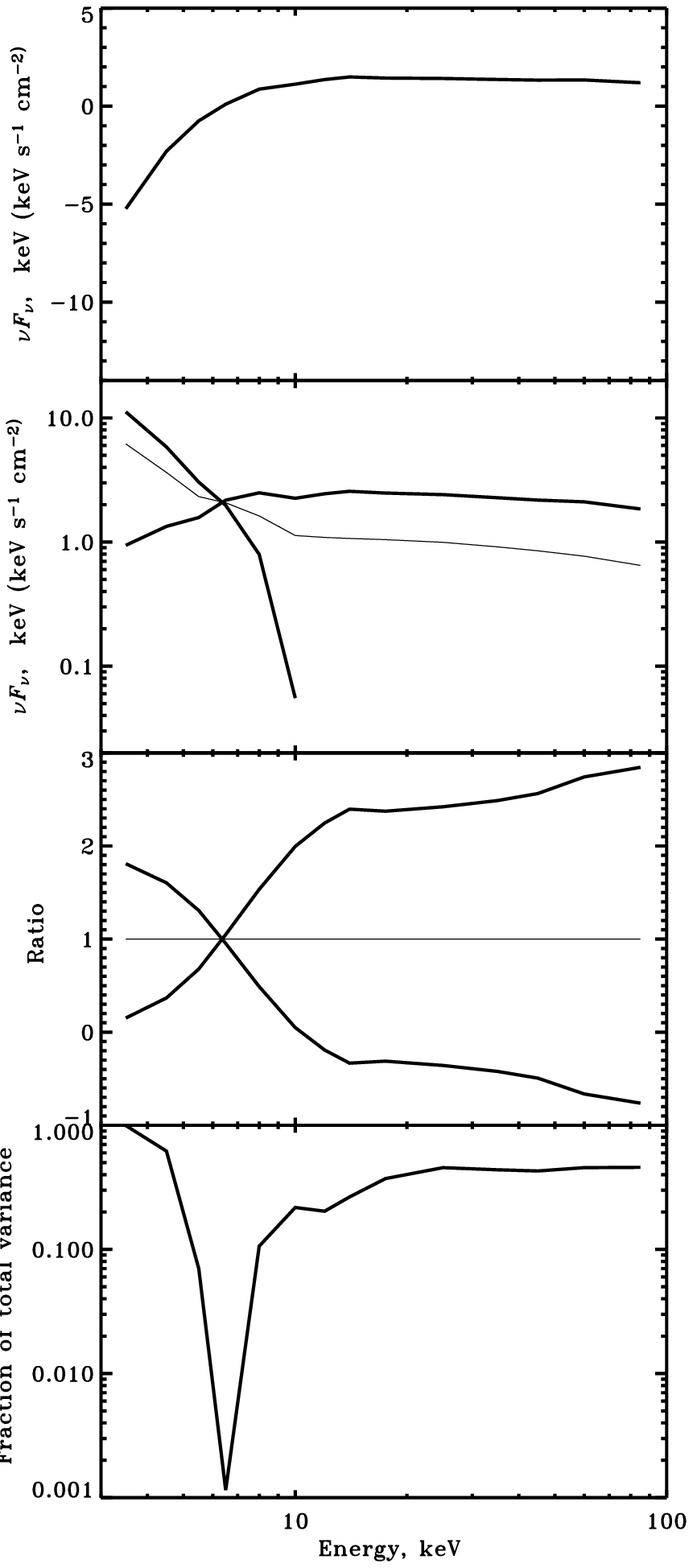}
\includegraphics[height=9.cm, width=4.6cm]{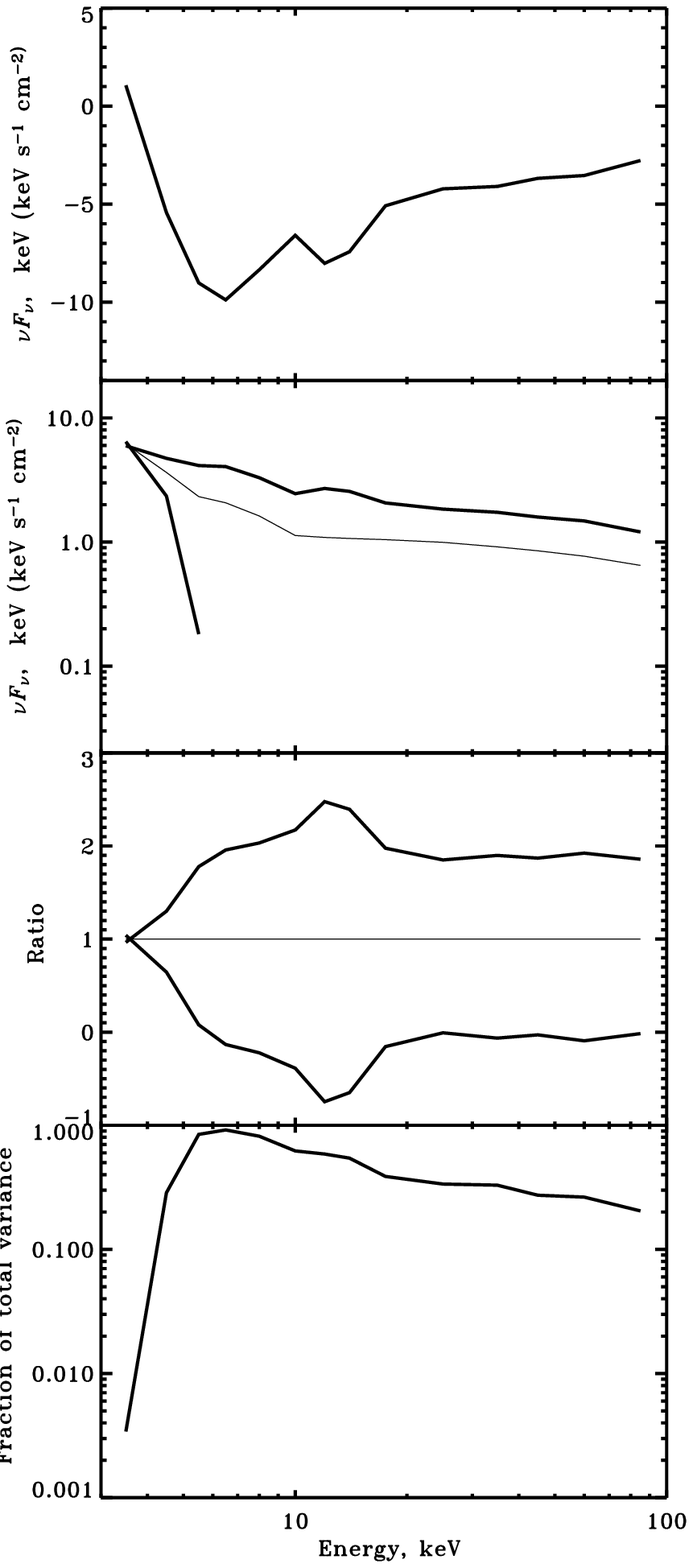}
\caption{PC1 ({\it left}) and PC2 ({\it right}). From the top panel: the shape of the 
principal component, the effect on the shape and normalisation of the spectrum, the ratios
maximum (and minimum) spectrum to the average one and the contribution to
the variance are shown. \label{fig:pca}}
\end{figure*}

\section{Results}
The hardness ratio 30-40 keV to 4-5 keV (Fig. \ref{fig:lc}, {\it bottom}) clearly shows the progressive softening of \mysou; 
the spectral transition to the SIMS state
occurred between MJD 53232 and MJD 53233 \cite{belloni06}.
In the HIMS the best-fit parameters have been found 
to be $\Gamma$ = 1.72$\pm$0.04, E$_{c}$~=~89$\pm 3$ keV and kT$_{diskbb}$~=~0.36$\pm 0.01$~keV (Fig. \ref{fig:spec}, {\it left}).
The Fe line has been frozen at 6.7 keV and may be an instrumental effect.
In Fig. \ref{fig:spec} ({\it right}) the SIMS energy spectrum ({\it spec2}) and the best-fit model
being $\Gamma$~=~2.2$\pm$0.01 and kT$_{diskbb}$~=~0.61$\pm 0.02$~keV are shown.  

Two principal components describing the \mysou~ variability have been found:  
PC1 responsible for 75\% of the variance and PC2 for 21\%.
In Fig. \ref{fig:pca} we present from the top to the bottom:
\begin{itemize}
\item[1] the shape of each PC which would be added or subtracted to the average spectrum in order to reproduce the variability;
\item[2] the effects of PC1 and PC2 on the shape and normalisation of the spectrum: time-averaged spectrum (light line) 
and spectra obtained for the maximum and minimum (solid lines) observed values of the normalisation parameter; 
\item[3] the ratio of the maximum and minimum spectra to the average one; 
\item[4] the contribution of each component to the total variance as a function energy.
\end{itemize}

\section{Discussion}
We have presented \integral~ observations of \mysou~, which covered part of the 2004 outburst. 
We followed the source spectral evolution through the Hard-Intermediate and Soft-Intermediate states, 
until a softer state. The variability of \mysou~ can be described by two independent modes: 
\begin{itemize}
\item{spectral evolution with the spectrum pivoting around 6 keV (PC1) contributing at 75\%. 
This pivoting mode may be interpreted as caused by changes in the soft cooling photons flux in the hot Comptonising plasma 
associated with an increase of temperature of the accretion disc.}
\item{intensity variation of the hard power-law component (almost constant slope) 
on top of constant soft component (PC2) contributing at 21\%. It would be interpreted to be 
due to magnetic flares occurring in the corona.}
\end{itemize}

Unlike \cyg~ \cite{malzac06}, the pivoting mode has been found to be more important in \mysou.
Reasonably, because \cyg~ was observed during an incomplete spectral transition hard-to-intermediate, 
while \mysou~ achieved finally a softer state.

\section{Acknowledgements}
This work has been supported by the Italian Space Agency grant I/R/046/04.
Based on observations with \integral, an ESA project with instruments and science data centre funded by ESA member states 
(especially the PI countries: Denmark, France, Germany, Italy, Switzerland, Spain), Czech Republic and Poland, 
and with participation of Russia and the USA. MDS thanks Angela Bazzano for precious suggestions and support.

\end{document}